# Title: Y Chromosomes of 40% Chinese Are Descendants of Three Neolithic Super-grandfathers

**Authors:** Shi Yan[1,2]*, Chuan-Chao Wang[1], Hong-Xiang Zheng[1], Wei Wang[2], Zhen-Dong Qin[1], Lan-Hai Wei[1], Yi Wang[1], Xue-Dong Pan[1], Wen-Qing Fu[1,4], Yun-Gang He[2], Li-Jun Xiong[4], Wen-Fei Jin[2], Shi-Lin Li[1], Yu An[1], Hui Li[1], Li Jin[1,2]*

**Affiliations**

[1]Ministry of Education Key Laboratory of Contemporary Anthropology and Center for Evolutionary Biology, School of Life Sciences and Institutes of Biomedical Sciences, Fudan University, Shanghai 200433, China.

[2]Chinese Academy of Sciences Key Laboratory of Computational Biology, CAS-MPG Partner Institute for Computational Biology, SIBS, CAS, Shanghai 200031, China.

[3]Epigenetics Laboratory, Institute of Biomedical Sciences, Fudan University, Shanghai 200032, China.

[4]Department of Genome Sciences, University of Washington, Seattle, Washington 98195, USA

*Correspondence to: L. J. (lijin.fudan@gmail.com) or S. Y. (yanshi@picb.ac.cn).

**Abstract**: Demographic change of human populations is one of the central questions for delving into the past of human beings. To identify major population expansions related to male lineages, we sequenced 78 East Asian Y chromosomes at 3.9 Mbp of the non-recombining region (NRY), discovered >4,000 new SNPs, and identified many new clades. The relative divergence dates can be estimated much more precisely using molecular clock. We found that all the Paleolithic divergences were binary; however, three strong star-like Neolithic expansions at ~6 kya (thousand years ago) (assuming a constant substitution rate of $1 \times 10^{-9}$ /bp/year) indicates that ~40% of modern Chinese are patrilineal descendants of only three super-grandfathers at that time. This observation suggests that the main patrilineal expansion in China occurred in the Neolithic Era and might be related to the development of agriculture.

**One Sentence Summary:** Analyses of human Y chromosome revealed three rapid expansions in the Neolithic Age in East Asia

Demographic change is one of the central questions in understanding human history, and strong population expansions may be linked to various events as climate changes, alteration of social structure, or technological innovations. The recent advent of next-generation sequencing technology enabled a systematic analysis of the population history using the information from the whole genome with less ascertainment bias, so we can re-assess how the various factors have influenced the human population size and structure [1,2]. Recent analyses of mitochondrial genomes revealed that the expansions of female lineages of East Asians [3] and those of Europeans [4] started before the Neolithic Era, contradictory to the hypothesis that the agricultural innovation constitutes the primary driving force of population expansions [5]. These observations prompted this study to investigate expansions of male lineages.

The Y chromosome contains the longest non-recombining region (~60 Mbp, in which ~10 Mbp is non-repetitive) in the human genome [6,7], making it an informative tool for reconstructing genetic relationship of human populations and paternal lineages, and dating important evolutionary and demographic events [8-11]. However, the sequencing data of Y chromosomes of human populations is insufficient and biased even for those of current 1000-genome project for which coverage on Y chromosome was low (on average <1.4× in East Asian samples) [12].

According to the phylogenetic tree of Y chromosome, all the modern males could be categorized into 20 major monophyletic or paraphyletic groups (referred to as A to T) and their subclades [13,14]. Nearly all the Y chromosomes outside Africa are derivative at the SNP M168 and belong to any of its three descendent super-haplogroups – DE, C, and F [9,10,15], strongly supporting the out-of-Africa theory. The time of the anatomically modern human's exodus from Africa has yielded inconsistent results ranging from 39 kya [16], 44 kya [10], 59 kya [17], 68.5 kya [18] to 57.0 – 74.6 kya [19].

To achieve sufficiently high coverage in the non-recombining regions of Y

chromosome (NRY) and an adequate representation of individual samples, we selected 110 males, encompassing the haplogroups O, C, D, N, and Q which are common in East Eurasians, as well as haplogroups J, G, and R which are common in West Eurasians (see Table S1), and sequenced their non-repetitive segments of NRY using a pooling-and-capturing strategy.

**Results**

Overall ~4,500 base substitutions were identified in all the samples from the whole Y chromosome, in which >4,300 SNPs that has not been publicly named before 2012 (ISOGG etc.). We designated each of these SNP a name beginning with 'F' (for Fudan University) (see Table S2). We obtained ~3.90 Mbp of sequences with appropriate quality (at least 1x coverage on >100 out of 110 samples), and identified ~3,600 SNPs in this region. A maximum parsimony phylogenetic tree of the 78 individuals with good coverage was reconstructed (Fig. 1 and Fig. S1), the topology of which is congruent with the existing tree of human Y chromosome [13,20]. The tree contained samples from haplogroups C, D, G, J, N, O, Q, and R, and thus represented all the three superhaplogroups out of Africa – C, DE and F. In addition to the known lineages, many new downstream lineages were revealed. All the earlier divergences were found to be bifurcations, except for three star-like structures, i.e. multiple lineages branching off from a single node, were observed under Haplogroup O3a-M324, indicating strong expansion events.

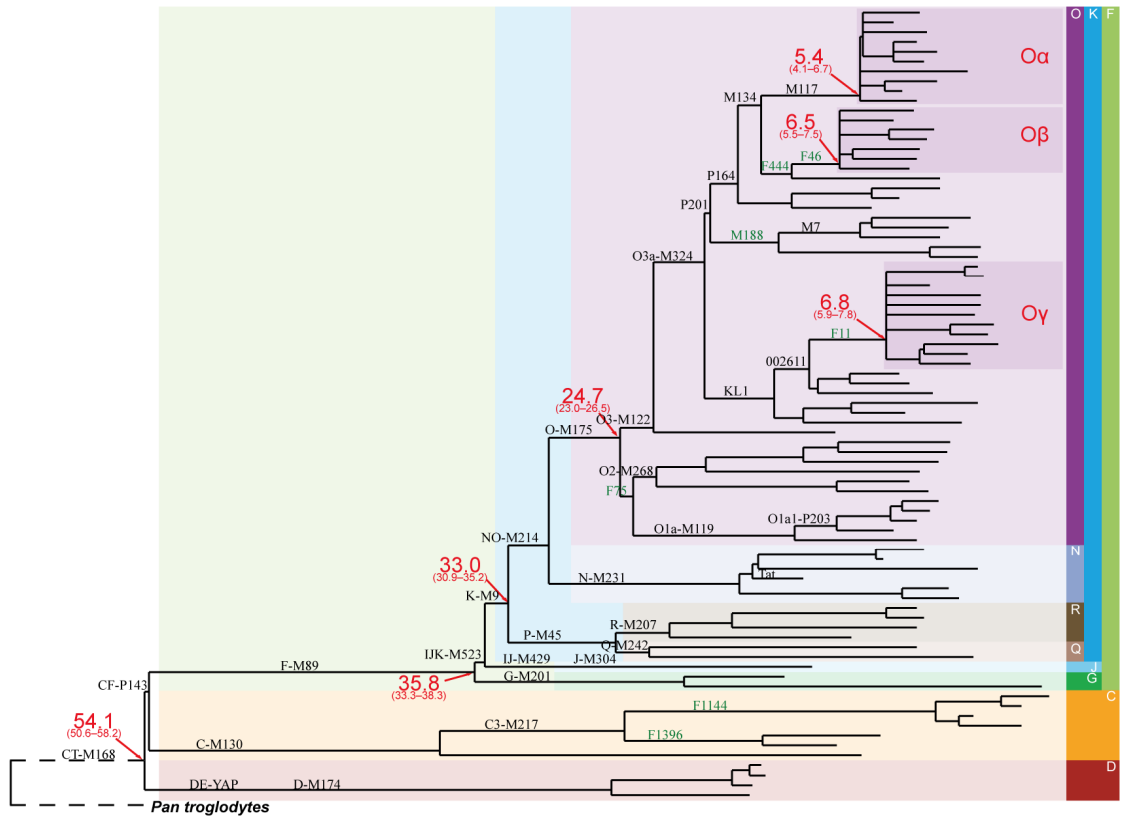

**Fig. 1. Phylogenetic tree of human Y chromosome, emphasizing the three star-like expansions (Oα, Oβ, Oγ).** The tree was constructed for 78 samples sequenced in this study, together with three published East-Asian genomes and a chimpanzee genome. The branch lengths (horizontal lines) are proportional to the number of SNPs on the branch. Numbers in red indicate the coalescence time (in years) and 95% confidence intervals of the node. For more details, see Fig. S1.

By using Bayesian method [21] with a constant mutation rate of $1 \times 10^{-9}$ substitution/base/year [7,19] (or one substitution per 256 years on 3.9 Mbp range), we calculated the date of each divergence event throughout the tree. The first divergence event out of Africa, i.e. between Haplogroup DE and the ancestor of C and F, is dated at 54.1 kya (95%CI 50.6 – 58.2), inside the range of previous estimations. Within the 3.9 Mbp range, only 3 SNPs were observed between the divergence events of DE/CF and C/F, indicating that DE, C, and F likely emerged subsequently in less than a thousand years. After diverged from Haplogroup C, no major split was observed in F for 18 thousand years, suggesting a strong bottleneck of F lineage. It should be noted that all the primary haplogroups (G, J, N, O, Q, and R) emerged before the last glacial maximum (LGM, ~20 kya), and most of the presently known East Eurasian clades have branched off in the late Upper Paleolithic Age (before 10 kya). Only binary divergences on this tree occurred before 7 kya, suggesting that during the Paleolithic Age, slow population growth and frequent bottlenecks eradicated most of the ever existing clades [22].

The most surprising discovery in the tree is the three star-like expansions in Haplogroup O3-M324, i.e. under the M117 clade, the M134xM117 paragroup, and the 002611 clade. Here we denote the three star-like expansions as Oα, Oβ, and Oγ, respectively (see Discussion). Since the sample selection for high-throughput sequencing was intended for representing a wide variety of clades in East Asian populations, a star-like expansion indicates successful expansion of male lineages within a very short period (<500 years). These three clades are present with high frequency across many extant East Asian populations [23,24] and encompass more than 40% of the present Han Chinese in total (estimated 16% for Oα, 11% for Oβ, and 14% for Oγ) [20]. It is conspicuous that roughly 300 million extant males are the patrilineal progenies of only three grandfathers in the late Neolithic Age.

The expansion dates are estimated 5.4 kya for Oα, 6.5 for Oβ, and 6.8 for Oγ (Fig. 1), after the shift to intensive agriculture in North China (since 6.8 kya) [25,26], in particular, during the Yangshao Culture (6.9 – 4.9 kya) in Central Yellow River Basin, Majiayao Culture (6.0 – 4.9 kya) in the Upper Yellow River Basin, and the Beixin (7.4 – 6.2 kya) – Dawenkou Culture (6.2 – 4.6 kya) in the Lower Yellow River Basin [27]. We therefore propose that in the late Neolithic Age, the three rapidly expanding clans established the founding patrilineal spectrum of the predecessors in East Asia. Since all the sequenced Han Chinese M117+ samples are under the Oα expansion, and M117+ subclade exists in moderate to very high frequency in many Tibeto-Burman ethnic groups [28-30], it would be of interest to know when the M117+ individuals in other ethnic groups diverged with the ones in Han Chinese, and whether they are also under the Oα expansion, in order to trace the origin and early history of Sino-Tibetan language family.

This study shows that all the strongly expanding Y chromosomal haplogroups (i.e. O-M175 or C-M130) had already migrated to East Asia more than 20 thousand years before their Neolithic expansion, thus supporting a boom of local farmers in China, which is consistent with the independent origin of agriculture [31], while differing from the case in Europe, where immigrant farmers from the Middle East contributed to the majority of modern Y chromosomes [32].

**Discussion**

Although most of the sequences in this study were obtained from individuals in China, the haplogroup representation (C, D, G, J, N, O, Q, and R) already enabled us to calculate the times of all the major divergence events out-of-Africa, like G/IJK, NO/P etc., since the times were achieved using the hypothesis of molecular clock, and the results of divergence time between haplogroups would not be affected by from whichever continent or country the individuals were sampled. One good sequence from each of two haplogroups is enough for calculating their divergence time, and more sequences could only help to enhance the precision but would not greatly change the result.

The significant improvement of accuracy of dating in this study comparing to former East Asian studies is attributed to the large number of newly discovered SNPs. It is noted that the relative standard deviation of calculated divergence time is in inverse proportion to the square root of observed SNP occurrence (see SI Discussion). Furthermore, the average counts of SNPs from the common ancestor of CF/DE to a modern individual is 210 in this study, limiting the theoretical 95%CI to only ±13.6%, comparing to 9 SNPs on average in the previous study with the 95%CI over ±60% [16]. Considering that 3.9 Mbp range constitutes only less than half of 10 Mbp non-repetitive region in Y chromosome [7], the time resolution of east Asian Y chromosome phylogeny is expected to be doubled in the near future.

The determination of mutation rate is a crucial question in calculation of the absolute divergent times, which caused the most dating differences among the studies. As revealed by previous studies, this inconsistency of mutation rate was resulted from two aspects: among different regions of the chromosome, and between older and younger time scales. The former has been disclosed in a study of autosomes, that the base substitution rate of CpG bases is 9.5-fold that of non-CpG bases [33], as well as for mitochondrion, the substitution rate was not only differentiated between coding and control regions, but also in a base-by-base manner [34]. It is worth to point out that recently, Wei et al. published a similar study about Y chromosome sequencing of 36 individuals (mainly Haplogroup R1b and E1b), in which 3.15 or 8.83 Mbp range was sequenced [19], and they achieved a time of out-of-Africa at 57 – 74 kya using various methods, which is slightly older than our result (54 kya), although the same mutation rate of $1 \times 10^{-9}$ substitution/base/year were employed. The difference could be ascribed to the regions chosen for date estimation; we compared the regions that Wei et al. and we studied, and found that in their study, the SNP density in the region that was sequenced only in their study is significantly higher than that in the region that both studies have sequenced (P<0.005) (Table S3).

The difference between long-term (evolutionary) and short-term (genealogical) mutation rates has also been observed before. For calculating the divergence time using Y-chromosomal STR (short tandem repeat), the father-son mutation rate is about three times the "evolutionary" [35]; similar rate difference was also observed for mitochondrial nucleotide substitution rate [34,36]. This controversy is usually explained by selection on deleterious mutations [37]. The autosomal genealogical substitution rate was estimated at $1.2 \times 10^{-8}$ substitution/base/generation [33,38], which is less than half of the rate we used in this study. However, due to that 80 – 85% of de novo mutations are attributed to the father's side [33], and that the Y chromosome contains the least genes among the chromosomes and thus underwent lessened purifying selection [39], the mutation rate used in this study is still compatible with the previous studies.

We also compared human and chimpanzee Y chromosomes (see SI Methods), and found ~45,800 substitutions between the two species which fall into the range that we compared for human samples; roughly 1/4 intra-human SNPs have no homologous loci on chimpanzee Y chromosome. Assuming the divergence between human and chimpanzee was at ~6,000 kya [40] and a constant substitution rate, the divergence time for DE and CF would be only 40 kya, which is younger than our result. This suggests that base-substitution rate between human and chimpanzee is higher than the rate inside human species, which can be explained with the huge interspecies difference of the Y-chromosome structures, and the observation that the chimpanzee Y chromosomal genes decayed faster than human [39].

To overcome the factors for uncertainty of mutation rate, a calibration with series of samples of comparable time scales might be used. For the case of mitochondrion, a recent study, in which several C-14 calibrated ancient complete sequences (4 – 40 kya) were incorporated into the tree, made the absolute dates much more convincing [41], and we expect a parallel calibration for the Y chromosome in the near future.

Despite of the mutation rate uncertainty, we evaluate our calculation of absolute divergence time as acceptable. Firstly, our out-of-Africa date (54.1 kya) is still within the range of previous estimations (39 – 74.6 kya). Secondly, the out-of-Africa date is similar to the recent estimation of two great mitochondrial expansions outside Africa – M (49.6 kya) and N (58.9 kya) [42]. Thirdly, it is not contradictory to the emergence of earliest modern human fossil out of Africa (e.g. ~ 50 kya in Australia) [43].

The accumulative substitution counts from the DE/CF divergence to a modern invidual varies from to 168 (YCH113) to 241 (YCH198). Despite of this variation, by testing the assumption of molecular clock for the tree, the null hypothesis of a molecular clock could not be rejected (P>0.05) using PAML package v4.4 [44] with the GTR model, unlike the mitochondrial tree from complete sequences, which showed violation to the clock assumption [42]. Part of the branch length variation may come from the false negative detection of SNPs, especially on a long terminal branch; however, this effect was mostly eliminated that we chose only the sequences with good quality for time estimation, so the branch length difference for these sequences should mainly reflects the real variation, and should have little effect in time estimation.

The current Y haplogroups were named according to the rule of Y Chromosome Consortium (YCC) [14]. Along with increasing clades being discovered, the present nomenclature became cumbersome in some cases, e.g. 'R1b1b2a1a2d3a' in the ISOGG tree 2010 (http://www.isogg.org), which is prone to frequent name changes and hard to remember. Another commonly used nomenclature such as 'O-M117' or 'O-F46' is also not suitable for a determined star-like expansion, since there are many SNPs found ancestral to the star point, while these SNPs may be found not all equivalent in the future, e.g. some individuals might be found M117+ but not belonging to the star expansion, then the name of the star point must be renewed. Therefore, here we propose a modification to the current nomenclature system: for any important star-like expansion that leads to large population (e.g. several millions) and multiple lineages (≥5) in short time as revealed by long-range sequencing (>1 Mbp was needed in order to limit the expansion within 1,000 years), a lineage name with lowercase Greek letter is applied directly after the Latin capital letter of the first-class haplogroup name. For example, the star-like expansions under M117, F46, and F11 are now named as Oα, Oβ, and Oγ, respectively, and their downstream lineages should still be named following the rule of YCC 2002, with Arabic number succeeding the Greek letter, e.g. Oα1a1. These names of the star-like expansions are not bound to any single defining SNP (e.g.

M117), but to the expansion itself, i.e. the expansion names should always keep unchanged despite new side clades would be found to its upstream, in order to keep the nomenclature stable. For the currently equivalent SNPs on the branch leading to the expansion, we will know the occurring order only after vast amount of samples being genotyped for those SNPs.

Since all the Paleolithic divergences of Y chromosome lineages are binary, the three roughly contemporaneous star-like expansions revealed in this study indicate a remarkable demographic change in the late Neolithic Age. The earliest agriculture in North China emerged before 10 kya [45], however, no distinct Y chromosomal expansion could be related to this event. The three star-like expansions happened several thousand years later, thus are likely linked to middle Neolithic cultures such as Yangshao (6.9 – 4.9 kya) and Dawenkou Culture (6.2 – 4.6 kya) in the Yellow River Basin [27]. During this period, agriculture became mature and intensive, and the majority of human diet shifted from food collection into production [46,47]. Crop harvest constituted a more stable food source than hunting and gathering, and enabled nourishing population at higher density. In addition, liberation of males from hazardous hunting might have enhanced male viability into adulthood, thus the effective population size of Y chromosome increased. Besides the progress in agriculture, changes in social structure might also contribute to the patrilineal expansion. In the middle and late phases of Yangshao and Dawenkou culture, the burial customs showed a gradual transition from an egalitarian matrilineal society into a hierarchical patrilineal one [48,49]. Interestingly, the major maternal expansions in China shown by mitochondrial tree (among which are also several star-shaped expansions) occurred much earlier, at the late Paleolithic Age [3]. This immense non-synchrony between maternal and paternal expansion suggests a possible transition of social structure, that in the late Neolithic Age, a few paternal lineages achieved greater advantage on the existing basis of the population that started expansion since the Paleolithic Age. After the strong Neolithic expansions, the reproductivity advantage of the farmers lasted for 4,000 years, until most of the gatherer-and-hunter tribes in the Yellow River Basin were absorbed by the farming societies of Huaxia, from which the Han ethnicity was formed.

Although without ancient DNA proofs, we cannot yet confirm the initial expanding regions of these three clans, whether they were original in the middle or lower reach of Yellow River Valley or migrated from the vicinity, we are now at least certain that a majority of Han Chinese did derive from just a few patrilineal ancestors in the Neolithic Age. Whether each of them could be related to the legendary Emperors *Yan* and *Huang* or their tribes, is to be solved with more prudence and with the help of interdisciplinary genetic, archeological, ethnical, and documentary studies.

**Methods:**

*Samples*

We collected whole blood from ~800 Chinese male volunteers with informed consent, under the approval of the Bioethics Committee of Fudan University. Genomic DNA was extracted using QIAamp DNA Blood Mini Kit (QIAGEN, Hilden, NRW, Germany). SNaPshot multiplex kit (ABI, Carlsbad, CA, US) was used for typing Y chromosomal SNPs according to the most recent phylogenetic tree [13,20], and 17 Y-STRs were determined with Y-filer kit (ABI, Carlsbad, CA, US). We selected 110 samples for next-generation sequencing, considering Y haplogroup, STR haplotype, as well as ethnic origin, in order to represent a broad spectrum of Y chromosome lineages of Chinese populations (Table S1). The selected samples covered most sublineages of Haplogroup O (72 samples), as well as Haplogroup C, D, G, J, N, Q, and R.

*Library preparation*

Genomic DNA of the selected samples were sheared using Bioruptor UCD-200 (Diagenode, Liège, Belgium) to 200 –250 bp length, then were fixed to blunt-end, added 3´-A tail, and ligated with barcode-linked Illumina paired-end adaptors (Table S1). Ligation products were amplified by PCR, and 300 – 350 bp sections were extracted through agarose gel electrophoresis. Except for one sample (YCH53), the others were pooled into 8 pools, with 10 – 15 samples in equal amount in each pool (Table S1). NRY was enriched using custom designed bait library (see below) of G3360-90000 SureSelect kit for Illumina paired-end (Agilent, Santa Clara, CA, US) (the baits were listed in Table S4). After another round of amplification, the pools went through single-end or paired-end sequencing with either GAIIx or HiSeq2000 sequencer for 100 or 2×100 cycles (Illumina, San Diego, CA, US).

*Bait design*

For Agilent SureSelect enrichment, bait library was designed with the following procedures: we first simulated reads mapping by generating 70-bp fragments of reference Y chromosome (hg18 or NCBI build36) (http://hgdownload.cse.ucsc.edu/downloads.html#human) for each 10 bp, e.g. chrY:1-70, chrY:11-80 etc. The fragments were then mapped on the complete hg18 genome using soap2 aligner (http://soap.genomics.org.cn/)[50]. All the match results with 0 – 2 mismatch bases on all chromosomes were summed up, and only the fragments without any repetitive matches (on either Y or other chromosome) were kept as unique fragments. The range of those unique fragments was combined, and the combined ranges that are at least 240 bp long were selected for bait design on Agilent eArray website (https://earray.chem.agilent.com/earray/). Totally 40,379 baits covering 4,292,864 bp were successfully designed and ordered for production. The ranges (on hg18) of the generated baits are listed in Table S4.

*Processing of next-generation sequencing data*

The barcodes were removed and the reads were assigned to each sample. For paired-end sequencing, the reads were assigned only when the both barcodes were the same. The reads were mapped to hg18 using *bwa* aligner (version 0.5.8)[51], and sam files were generated. Reads that were uniquely mapped on Y chromosome were extracted and transformed into bam file with *samtools* (version 0.1.8)[52]. Duplicates were removed by either Picard's *MarkDuplicate* (http://picard.sourceforge.net) (for single-end) or *samtools rmdup* (for paired-end). Indels were re-aligned using GATK [53,54], and after *samtools mpileup*, variations were called under the following criteria: for one sample, the position where the alternative allele (compared to hg18) must be ≥

2× coverage and at the same time ≥ 3/4 of total coverage. All the variance candidates were collected, and were genotyped on all the sequenced samples. Out of those candidates, SNPs were semi-manually filtered considering consistency to the Y chromosomal phylogeny, coverage (especially for the private SNPs, a minimum of 4× was required), and flanking sequence (to avoid those included or next to a homopolymer or an STR). Three other publicly available East Asian genomes, YanHuang (YH) (O1a1-P203) [55], KoRef (SJK) (O2b-M176) [56], and GMIAK1 (O3a2c*-P164xM134) [57] were also included in analysis.

*Time estimation of the nodes in the phylogenetic tree*

A coverage filter was applied for time estimation, i.e., only the loci with good coverage among the sequenced samples, i.e., with more than 100 out of 110 genotyping results with an unambiguous 0 or 1 were selected for phylogenetic reconstruction (0 for same as reference, 1 for mutation, question mark "?" when neither reference or alternative counts for more than 3/4 for this sample, and a minus mark "-" for no coverage, see Table S5). SNPs were extracted into pseudo-sequences, and a maximum parsimony tree of only good- and moderate-quality sequences was calculated using ARB program [58].

To avoid uncertainty in downstream branches that might influence branch lengths, we used only good-quality sequences for Bayesian time estimation (with on average >6× coverage at targeted regions and no obvious mislabeling, see Table S1). We used BEAST [21] for calculating the divergence time of each node in the phylogenetic tree. All the 47 high-quality sequences together with YH and SJK were used for time estimation. Pseudo-sequences were generated from SNPs, and appropriate DNA substitution model was determined with MrModeltest 2.3 [59] for subsequent Bayesian MCMC analysis. For Bayesian MCMC analysis, the times of each cluster were estimated using BEAST1.6.1 [21,60]. Each MCMC sample was based on a run of 20 million generations sampled every 10,000 steps with the first 2 million generations regarded as burn-in. To test the assumption of molecular clock for the tree, we used PAML package v4.4 with the GTR model. The null hypothesis of a molecular clock cannot be rejected (P>0.05) by comparison between the models. We used the GTR model of nucleotide substitution determined with MrModeltest 2.3 with a strict clock. The single nucleotide substitution rate was set as $1 \times 10^{-9}$/nucleotide/year. The effective sample size of the coalescent prior was above 900. A relaxed clock was also employed for comparison and the results were similar.

Chimpanzee genome (panTro3)[61] (http://hgdownload.cse.ucsc.edu/downloads.html#chimp) was used as comparison for time estimation. All the base substitutions of chimpanzee genome comparing to hg18 were discovered using the similar method as for bait design: the simulated 100-bp-long reads at each 10 bp were generated and mapped onto hg18 using *bwa*, and SNPs were discovered. The SNPs intra human beings were also genotyped for the chimp reads. The root for the human samples in this study was thus determined.

**Acknowledgments:**

This study was supported by the National Science Foundation of China (NSFC) grant 30890034 and 31071096.

**Author Contributions:**

S. Y. and L. J. designed the study, analysed data and wrote paper; S. Y., C.-C. W. and Z.-D. Q. performed the next-generation sequencing experiment; W. W., H.-X. Z., Y. W., X.-D. P., W.-Q. F., W.-F. J. and Y.-G. H. contributed to next generation data analyses, statistics and modelling; L.-H. W. conducted geographical analysis and contributed to the maps; L.-J. X. helped the experiment design; S.-L. L. and Y. A. coordinated experimental resources; H. L. was involved in discussion.

**Competing Financial Interests statement:**

The authors declare no competing financial interests.

# Supplementary Materials:

**Additional Discussions:**

*Precision in observing SNP occurrence*

Assuming the base substitution rate is constant, and each mutation is an independent random event, the observed occurrence of mutations from an ancestor to the descendant should follow Poisson distribution with expectation $\lambda$, which can be approximated as the observed mean of mutation occurrence. The standard deviation $\sigma$ of Poisson distribution equals to $\sqrt{\lambda}$. When $\lambda$ is a large number, the Poisson distribution can be simplified as a normal distribution $X \sim N(\lambda, \lambda)$. Thus the relative range of 95%CI is roughly $\pm \frac{2\sigma}{\lambda} = \pm \frac{2\sqrt{\lambda}}{\lambda} = \pm \frac{2}{\sqrt{\lambda}}$, i.e. in inverse proportion to the square root of expectation $\lambda$. In this study, we calculated the mutation rate with 19 independent branches, within each branch there are averagely more than 20 SNPs, therefore our rate is much more accurate than that from the previous study [7], in which only 4 SNP were identified. We also predict that when the whole NRY (~10 Mbp, with some regions prone to recurrent mutations) of these samples is sequenced in the future, the discovered SNPs would be roughly doubled, and the precision would be enhanced by 1.4×, i.e. the 95%CI would be reduced from ±13.6% to less than ±10%.

*Estimation of captured range*

Considering both the positions on the designed bait regions and the 250 bp flanking regions by the baits, the range designed to be captured were 7.017 Mbp. Out of all the reads, 32 – 39% were mapped on this range (fluctuating among each pool). Since the size of whole human genome is about 3 Gbp and Y chromosome is haploid, the enrichment efficiency was about 2700 – 3300 times.

*Estimation of false positive and false negative rates of newly discovered SNPs*

We tested 36 newly discovered SNPs under Haplogroup N or the 002611+ clade, all of which were shared by more than one tested samples and conform to the phylogenetic tree. All these SNPs were validated by Sanger sequencing, and are thus highly reliable. We also tested 19 new private SNPs (found derivative in only one sample with at least 4× coverage), all but one were validated by Sanger sequencing. The false private SNPs could be due to any of sequencing error, mismatch during PCR steps, mapping error to homologous sequence, or misalignment near indels.

False negatives, i.e., undiscovered SNPs inside the sequenced 3.9 Mbp region can also be a factor that affects divergence time calculation. Specifically, most of those false negatives may exist in the terminal branches, or the private SNPs, due to low sequencing depth. Our procedure should have lowered this effect. Firstly, only the 47 individuals with averagely >6× depth and without clear traces of contamination (in bold italic in Fig. 1 and Fig. S1) were used for time estimation, and other 31 sequences with moderate quality (less depth or with slight traces of confusion) were kept in the tree in delivering more phylogenetic information but not used for time estimation. Due to the fact that Y chromosome is haploidic, for the non-repeated regions of Y chromosome, under a condition of an error rate of 0.1% (Illumina), 2× sequencing depth should be enough to discover a new mutation, and 1× is enough for genotyping. Therefore the criterion we used (on average >6× after removing duplicates) should not cause many false negatives. Secondly, the individual with best sequencing quality, YCH53, with 239× depth sequenced in a separate lane, did not show significantly longer branch than other individuals, indicating that 6× depth should be enough to avoid false negatives. In fact, most problems of false negatives were attributed to intermingling between samples due to the sequencing quality of barcodes, and thus private SNPs showed heterozygous and could sometimes not be effectively discovered. However, those individuals with any traces of intermingling were excluded from time estimations.

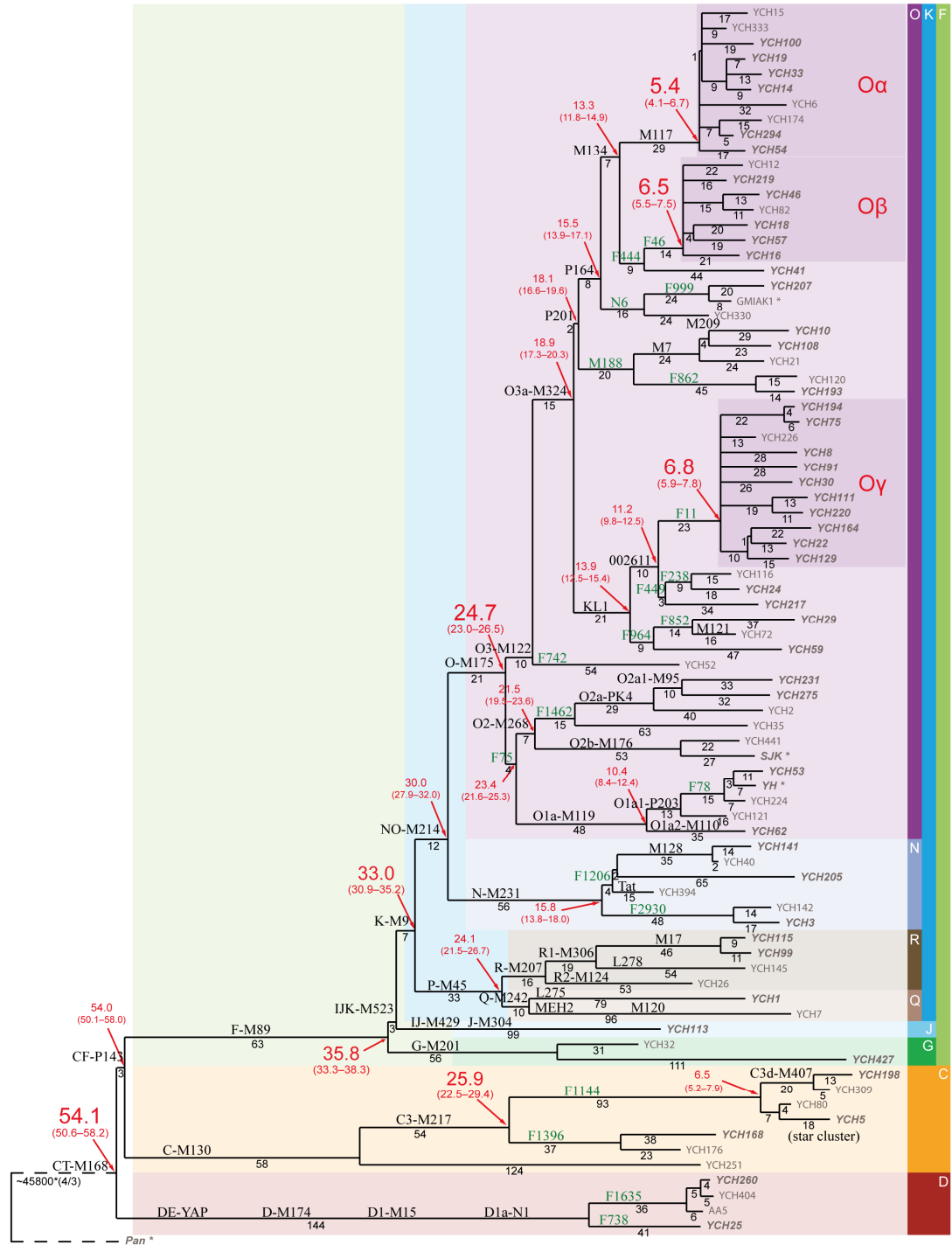

**Fig. S1. Phylogenetic tree of human Y chromosome.** The tree is constructed for 78 samples sequenced in this study, together with three published East-Asian genomes (YH, SJK, GMIAK1) and a chimpanzee genome (Pan), which are labeled with '*'. Except for YCH145 (Spanish), SJK and GMIAK1 (both Korean), all the human samples are Chinese. The branch lengths (horizontal lines) are proportional to the number of SNPs on the branch, and the SNP numbers are labeled under the branches). The SNPs labeled on the horizontal lines are only representative. The SNPs labeled in green represent newly recognized clades in this study. The estimated coalescence time (in years) for the nodes are calculated only from good-quality (> 6× coverage) human

sequences (in bold italic) by BEAST with relaxed clock (see SI Methods), and the numbers in brackets are for 95% confidence intervals (ignoring uncertainty in mutation rate).

*Bottleneck and expansion viewed from the tree*

The time and shape of divergences in the phylogenetic tree along with the information of extant geographic distribution may provide information on when, where and how the haplogroups resided and developed. Whether an ancient expansion would be observed today depends not only on how strong the expansion was, but also on how the descendants survived in the later periods. Since the expansion of a population is an intrinsic continuous property, the final outcome of an expansion depends on the environment. A line segment on the tree without divergence during a long time may indicate a strong bottleneck occurred at the later end of the segment. On the other hand, a strong expansion during a short time does not simply mean that the period is extraordinary suitable for this clade, but rather that since this expansion, no prevalent and vital bottleneck has happened to counteract this expansion. Therefore, the stable food supply that was profited from agriculture likely provided the prerequisite of the successful star-like expansions. The time and shape of a divergence event may indicate whether the environment befits the survival. For example, frequent splits on a certain clade during the LGM period may indicate a southern (warmer) location, while an expansion that only happened several thousand years later than the LGM may indicate a relatively northern existence.

*A refined history of the haplogroups*

According to the divergence times calculated from the tree (Fig. S1) and the present geographical distribution of each haplogroup, a renewed migration route of East Eurasian haplogroups was proposed (Fig. S2 a – d). The base map was generated using the GMT Tools (http://gmt.soest.hawaii.edu/).

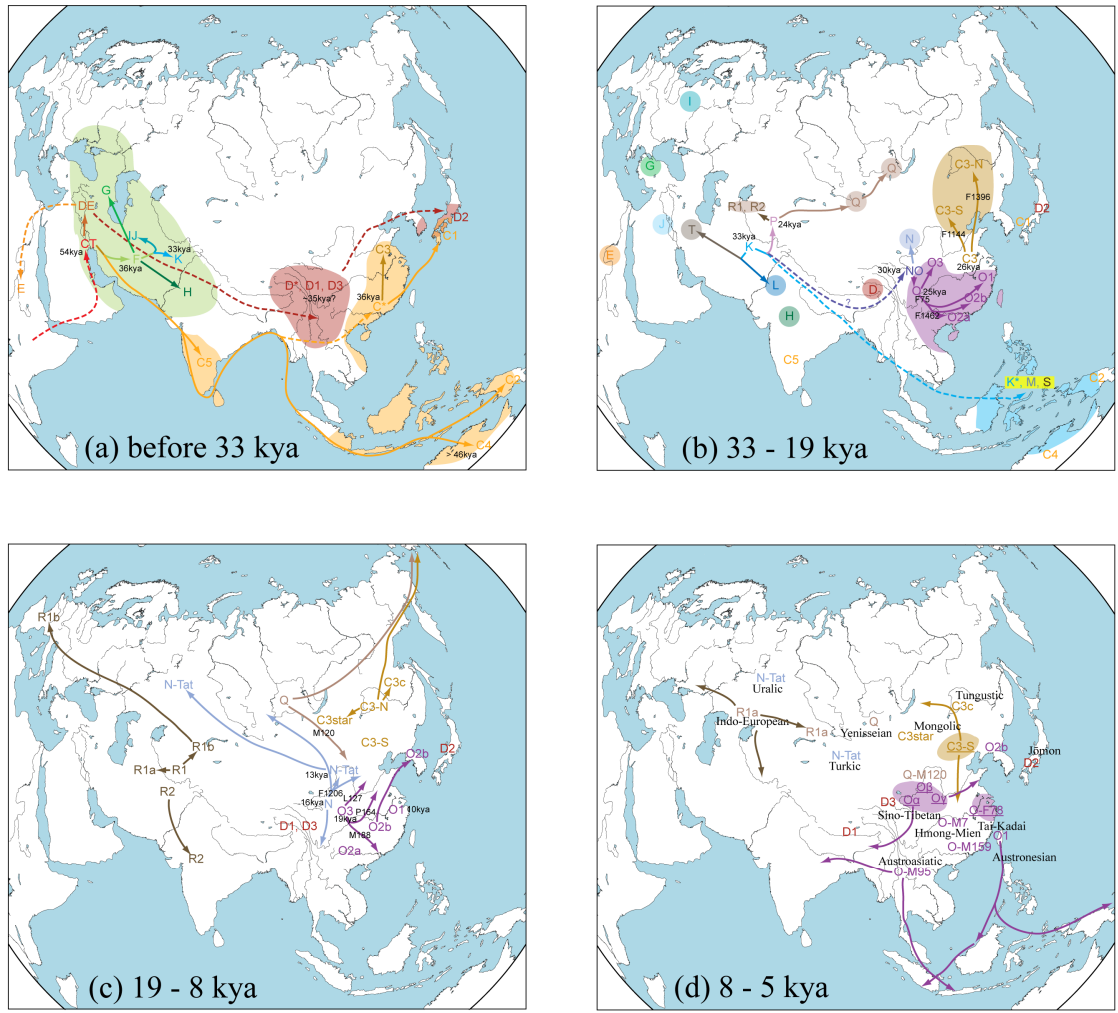

**Fig. S2. Revised migration routes of modern human.** (a) Split of the first out-of-Africa ancestor and early migration in Asia. (b) The emergence of the main haplogroups before and during the LGM. (c) Foundation of present haplogroup distribution before 8 kya. (d) Major population expansion events in East Asia (shaded) in the Neolithic Age and their probable relationship with modern language families.

It remained mysterious that how many times the anatomically modern human migrated out of Africa, since that among the three superhaplogrous C, DE and F, Haplogroup F distributes in whole Eurasia, C in Asia and Austronesia, D exclusively in Asia, while D's brother clade E distribute mainly in Africa [62], so there are two hypotheses, 1) haplogroups D and CF migrated out of Africa separately; 2) the single common ancestor of CF and DE migrated out of Africa followed by a back-migration of E to Africa. From this study, the short interval between CF/DE and C/F divergences weakens the possibility of multiple independent migrations (CF, D, and

DE*) out of Africa, and thus supports the latter hypothesis [63] (Fig. S2 a).

Haplogroup D is comprised of subclade D1-M15 and D3-P99, both in continental East Asian, especially frequent in Tibet [63], subclade D2-M55, nearly exclusively in Japan, and paragroup D*, which was discovered mainly in Tibet as well as on the Andaman Islands. In this study, only D1 and D3 samples were sequenced. The divergence time of D1 and D3 was at ~36 kya. Except for the sample YCH177 (Zhuang ethnicity), all the tested D1 samples (Han and Yi ethnicities) are derivative at SNP N1 [64].

Together with Haplogroup D, C is also considered as one of the harbingers in East Eurasia and Australia [65]. Soon after its divergence with F, Haplogroup C moved eastwards along the coast of Indian Ocean, reached India and China, and might be associated with the earliest known modern human inhabitants in Australia at 46 kya [66]. In China, the vast majority of Haplogroup C belongs to C3-M217 [67], which constitutes ~10% of Han Chinese, as well as great part of Altaic-speaking populations, e.g. Mongol, Manchu, and Kazakh. Here we identified two clades of C3 which split at 25.9 kya: a northern clade (C3-n) with SNP F1396, including a Mongol and a Manchu sample, and a southern clade (C3-s) with SNP F1144, including all sequenced Han Chinese C3 samples. The STRs of YCH168 (Mongol ethnicity) is close to the 'star-cluster', which is abundant in the steppe ethnicities [68], indicating that a substantial part of Altaic-speaking population belongs to C3-n. The southern clade expands rather late (only about 6.5 kya, i.e. in the Neolithic Age), including most former C3* individuals in Han Chinese. Interestingly, the subclade C3d-M407, which is common in Sojot (Turkic) and Buryat (Mongolic)[69], originated only after this expansion of C3-s. The C3-s clade showed a similar expansion time comparing to the three star-like expansions under O3, and probably will also be found a multifurcation, if more samples will be sequenced.

The Superclade F did not undergo major split since 54.0 kya, until the divergence of Haplogroup G and IJK at 35.8 kya, which was followed by the emergence of all major haplogroups (IJ, and K, and its subclades NO, P, and LT) during the following 3,000 years. Considering the present distributions of these haplogroups, we suggest that the expansion occurred in Central or South Asia. Since 30 kya, the Eurasians became adapted to colder environment and thrived in more northern areas like Europe and North China (Fig. S2 b), despite of the coming last glacial maximum (LGM, around 20 kya). Haplogroup NO migrated to East Asia and split into N and O at 30.0 kya.

Haplogroup P diverged into Q and R at ~24.1 kya, slightly before the LGM. Most Q individuals in Han Chinese belong to the Q1a1-M120 clade, while R's in Han Chinese are mostly R1a1-M17. The separation events of R1 and R2, and R1a and R1b are estimated here at 19.9 and 14.8 kya, respectively. R1b roamed till the Atlantic coast, forming some of the non-Indo-European groups (e.g. Basque)[32].

This study leads to a discovery of 265 new SNPs under Haplogroup N-M231, adding significantly to the only 11 currently known SNPs [13]. Haplogroup N is frequently found in Tibeto-Burman [70,71], Austroasiatic [72], Altaic [73], Uralic [74], Slavic [75,76], and Baltic peoples [77,78]. Haplogroup N went through a bottleneck lasting for 14 thousand years (30 – 15.8 kya). Considering the last glacial period at 26.5 – 19 kya, when living space for human was probably very limited in the northern part of Asia, we therefore assume that at the LGM, the Haplogroup N survived in an area northern to Haplogroup O, e.g. the northern part of China. The expansion of Haplogroup N occurred at 16 – 13 kya, during which N1c-M46 and N1a-M128 separated. The former defining SNP of Haplogroup N1, LLY22g (double copied, CC > CA) is proved to be prone to recurrent mutation, since the sample YCH142, downstream of F842 clade, has CC genotype at LLY22g, while the other sequenced N samples have CA. The AA genotype was also observed in an N sample (genotyped in this study but not sequenced by next-generation method). It seems that CC and AA genotypes under

Haplogroup N were results of homologous recombination of the two copies. Therefore, we suggest to remove the SNP LLY22g from the current phylogenetic tree as the case of SNP P25, which has 3 copies on Y chromosome [79], and the phylogeny of Haplogroup N is to be rearranged using the single copied SNPs discovered in this study.

Haplogroup O, which covered 1/4 of all males on the world today [20], began frequent splitting into subclades before the LGM. The ancestor of O-M175 suffered an intermediate bottleneck event at 30 – 25 kya, and expanded rapidly at 24.7 – 21.5 kya, indicating a southern distribution during the LGM. We found that Haplogroups O1 and O2 share 6 SNPs (e.g. F75), forming a monophyletic lineage before joined with O3-M122 (the current nomenclature is not changed in this study). O1a1-P203 is the major clade of Haplogroup O1 in China, especially frequent (>20%) in the eastern provinces like Zhejiang and Jiangsu [20], corresponding to the Neolithic expansion of the ancient Yue (ancestral group of present Tai-Kadai and southern Han Chinese) [80]. The major expansion under P203 in Han Chinese population (F78 clade) occurred at about 5 kya, younger than the three star-like expansions in O3, fitting to the rice-growers in Liangzhu culture (5.3 – 4.0 kya). Since Tai-Kadai O1 samples were not included in this study, their divergence time with the East China Yue population is not yet clear.

Among the three main branches of Haplogroup O, O2 clade expanded the earliest, fitting the current distribution which is more at the south. All the sequenced O2-M268 samples other than O2b-M176 form a monophyletic clade, labeled by F1462, and the SNP PK4 lies inside this clade. Further genotyping of the newly discovered SNPs under F1462 clade will unveil the origin and migration routes and time of the Austro-Asiatic and Tai-Kadai peoples in South China, Southeast Asia and India [81].

Haplogroup O3 covers more than half of all the Han Chinese population [20]. Except for a few O3*-M122(xM324) individuals (sharing the SNP F742), all the sequenced O3 samples are found belonging to any of the three clades KL1+, M188+, or P164+, which diverged at ~18 kya. The SNP M188, which was defined as downstream of M7 [13], is now found upstream of M7. Although no M159-derived samples were sequenced in this study, it has been revealed that M188 is also upstream of M159 [82]. Therefore, M7 and M159 should be parallel and under M188, which is fraternal to P164 and under P201. Under the P164 clade, N6 [64] defines a monophyletic clade parallel to M134. It would be interesting to know the detailed subclade of the Austronesian P201*(xM134,M7) clade [23,83], in order to know their relationship to the continental peoples. All the sequenced M134xM117 samples also showed a monophyletic group with derived allele at F444. There are also secondary bifurcations under L127 and IMS-JST002611 (in short, 002611).

**Supplementary References:**